\documentclass[final,journal]{IEEEtran}

\usepackage[utf8]{inputenc}
\usepackage{booktabs}
\usepackage{cite}
\usepackage{listings}
\usepackage{color, colortbl}
\usepackage{graphicx}
\usepackage{svg}
\usepackage{epstopdf}
\usepackage{caption}
\usepackage{amsmath,empheq,amssymb,mathtools}
\usepackage{gensymb}
\usepackage{subfloat}
\usepackage{subfig}
\usepackage{siunitx}
\usepackage[shortlabels]{enumitem}
\usepackage{mathtools,cancel}
\usepackage[export]{adjustbox}
\newtheorem{theorem}{Theorem}[section]

\begin{document}

\title{`Constant in gain Lead in phase' element - Application in precision motion control}	
\author
{\IEEEauthorblockN{Niranjan Saikumar, Rahul~Kumar~Sinha, S. Hassan HosseinNia\\}
	\IEEEauthorblockA{Precision and Microsystems Engineering, \\
	Faculty of Mechanical Engineering, TU Delft, The Netherlands\\
	}
}

\maketitle
	
\begin{abstract}
	
This work presents a novel `Constant in gain Lead in phase' (CgLp) element using nonlinear reset technique. PID is the industrial workhorse even to this day in high-tech precision positioning applications. However, Bode's gain phase relationship and waterbed effect fundamentally limit performance of PID and other linear controllers. This paper presents CgLp as a controlled nonlinear element which can be introduced within the framework of PID allowing for wide applicability and overcoming linear control limitations. Design of CgLp with generalized first order reset element (GFORE) and generalized second order reset element (GSORE) (introduced in this work) is presented using describing function analysis. A more detailed analysis of reset elements in frequency domain compared to existing literature is first carried out for this purpose. Finally, CgLp is integrated with PID and tested on one of the DOFs of a planar precision positioning stage. Performance improvement is shown in terms of tracking, steady-state precision and bandwidth. 

\end{abstract}

	
\IEEEpeerreviewmaketitle
	
\section{Introduction}

\IEEEPARstart{P}{}ID continues to be popular in the industry due to its wide applicability, simplicity and ease of design and implementation. PID is used in high-tech applications from wafer scanners for production of integrated circuits and solar cells to atomic force microscopes for high-resolution scanning. With well-designed mechanisms and feed-forward techniques, high precision, bandwidth and robustness are being achieved. PID also lends itself to industry standard loop shaping technique for designing control using frequency response function obtained from the plant. However, the constantly growing demands on precision and bandwidth are pushing PID to its limits. PID being a linear controller suffers from fundamental limitations of Bode's gain phase relationship and waterbed effect \cite{tan2007precision,schmidt2014design}. It is self-evident that these can only be overcome using nonlinear techniques. However, most nonlinear techniques in literature presented for precision control \cite{kuo2005modeling,mittal1997precision,shan2002ultra,kuo2003large} are more complicated to design and/or implement and do not fit within techniques like loop shaping which are popular and widely used in the industry.

Reset control is a nonlinear technique which has gained popularity over the years and has the advantage of fitting within the framework of PID for improved performance. Reset involves the resetting of a subset of controller states when a reset condition is met. Reset was first introduced by J C Clegg in \cite{clegg1958nonlinear} for integrators to improve performance. Advantage of reset is seen in reduced phase lag compared to its linear counterpart \cite{krishnan1974synthesis}. This work has been extended over the years with other reset elements from First Order Reset Element (FORE) \cite{horowitz1975non}, Generalized FORE (GFORE) \cite{guo2009frequency} and finally to Second Order Reset Element (SORE) \cite{hazeleger2016second} introduced and used in control applications. Significant work can be found in literature showing the advantages of reset control \cite{beker1999stability,chen2001analysis,zheng2000experimental,hosseinnnia2013basic,hosseinnia2013fractional,hosseinnia2014general,heertjes2015design,beker2004fundamental,beker2001plant,wu2007reset}. However, in most of these cases, reset control has mainly been used for it's phase lag reduction advantage. Some works exist where reset has been used for phase compensation. In \cite{li2005nonlinear}, Ying et al. use the reset element to overcome the waterbed effect through mid frequency disturbance rejection by lowering the sensitivity peak.  In this case, reset is used to achieve a narrowband phase compensator, hence improving phase margin and performance. This compensator was further modified for improved performance and phase compensation in \cite{li2011reset} allowing for the use of notch for disturbance rejection without affecting stability margins. Reset control with optimized resetting action for improved performance has also been presented in \cite{li2011optimal}.

In this work, we present a novel reset element termed `Constant in gain Lead in phase (CgLp)' element which extends the use of reset to be used for broadband phase compensation. The element is designed using describing function analysis to work well within existing framework of PID, thus achieving industry compatibility. Improvement in precision and tracking is shown on a precision positioning stage. In Section \ref{prelim}, basics of reset systems are provided along with the definitions of reset elements present in literature. The novel GSORE element is presented in Section \ref{Freqbehav}. Further, while reset elements have mainly been analysed for their phase lag reduction in literature, other properties of generalized reset elements in frequency domain critical to CgLp design are discussed. Design and analysis of CgLp are presented in Section \ref{CgLpsec} followed by the inclusion of CgLp within framework of PID for broadband phase compensation. The application of this modified CgLp-PID controller on a precision positioning stage is dealt with in Section \ref{Application} to show improvement in performance. The conclusions and future work are provided in Section \ref{Concl}.						
		
\section{Preliminaries}
\label{prelim}

\subsection{Definition of Reset control}

A general reset controller can be defined using the following differential inclusions:

\begin{equation}
\Sigma_R = 
\begin{cases}
\dot{x_r}(t) = A_rx_r(t) + B_re(t) & \text{if } e(t) \neq 0\\
x_r(t^+) = A_\rho x_r(t) & \text{if } e(t) = 0\\
u(t) = C_rx_r(t) + D_re(t)
\end{cases}
\label{eq:reset}
\end{equation}

where $A_r$, $B_r$, $C_r$, $D_r$ are state-space matrices of the base linear system, $A_\rho$ is reset matrix determining the state after reset values. $e(t)$ is the error signal fed to the controller and $u(t)$ is the output of controller which is used as control input for plant. While other forms of reset like reset band and fixed instant reset exist in literature, the form provided above is the most popular, widely applied and tested. The reset controller of Eqn. \ref{eq:reset} generally consists of both linear and nonlinear reset part. The $A_\rho$ matrix is defined to reset only the appropriate states of controller.

\subsection{Describing function}
\label{resetprops}

The nonlinearity of reset elements creates the problem of designing controllers in frequency domain especially using industry popular loop shaping technique which uses Bode, Nyquist and Nichols plots. In literature, sinusoidal input describing function analysis has been used to analyse reset elements in frequency domain. In fact, the phase lag reduction advantage was seen by Clegg in 1958 using this technique. Although describing function does not accurately capture all the frequency domain aspects of reset, it is useful in providing necessary information for design and analysis. 

The describing function of generic reset systems as defined by Eqn. \ref{eq:reset} is provided in \cite{guo2009frequency} and this is used to obtain understanding of the system in frequency domain. The sinusoidal input describing function is obtained as 
\begin{equation}
G(j\omega)=C_r^T(j\omega I-A_r)^{-1}(I+j\Theta_\rho(\omega))B_r + D_r
\label{eq:tf}
\end{equation}
where
$$
\Theta_\rho  =  \frac{2}{\pi}(I + e^{\frac{\pi A_r}{\omega}})\Big(\frac{I - A_\rho}{I + A_\rho e^{\frac{\pi A_r}{\omega}}}\Big)\Big(\Big(\frac{A_r}{\omega}\Big)^2 + I\Big)^{-1}
$$

\subsection{Stability of reset elements and systems}
\label{resetstab}

Stability conditions given in \cite{banos2011reset} can be used to check closed-loop stability of reset control systems for SISO plants. The following condition has to be satisfied for ensuring quadratic stability:

\begin{theorem}	
	There exists a constant $\beta \in \Re^{n_r\times 1}$ and positive definite matrix $P_\rho \in \Re^{n_r\times n_r}$, such that the restricted Lyapunov equation
	\begin{eqnarray}
	P > 0,\ A_{cl}^TP + PA_{cl} < 0\\B_0^TP = C_0
	\end{eqnarray}
	has a solution for $P$, where $C_0$ and $B_0$ are defined by
	\begin{align}
	C_0=\left[\begin{array}{ccc}
	\beta C_{p} & 0_{n_r \times n_{nr}} & P_\rho
	\end{array}\right] , & &  B_0=\left[\begin{array}{c}
	0_{n_{p} \times n_{r}}\\
	0_{n_{nr} \times n_{r}}\\
	I_{n_r}
	\end{array}\right]
	\end{align}
	$A_{cl}$ is the closed loop matrix A-matrix
	\begin{equation}
	A_{cl} =\left[\begin{array}{cc}
	A_p & B_p C_r\\
	-B_rC_p & A_r
	\end{array}\right] 
	\end{equation}
	in which $(A_r,B_r,C_r,D_r)$ are the state space matrices of the controller defined by Eqn. \ref{eq:reset} with $n_r$ being the number of states being reset and $n_{nr}$ being the number of non-resetting states. 
	$(A_p,B_p,C_p,D_p)$ are the state space matrices of the plant.

\subsection{Reset elements}

The reset part of controllers defined by Eqn. \ref{eq:reset} have been presented as different reset elements in literature.

\subsubsection{Clegg Integrator (CI)}

Clegg or Reset integrator is the first introduction of reset technique in literature \cite{clegg1958nonlinear}. The action of resetting integrator output to zero when input crosses zero results in favoured behaviour of reducing phase lag from $90^\circ$ to $38.1^\circ$. CI is the most extensively studied and applied reset element in literature due to advantages seen in reduced overshoot and increased phase margins.

The matrices of CI for Eqn. \ref{eq:reset} are
\begin{equation*}
A_r = 0,\ B_r = 1,\ C_r = 1,\ D_r = 0,\ A_\rho = 0
\end{equation*}

\subsubsection{First Order Reset Element - FORE and its generalization}

CI was extended to a first order element as FORE by Horowitz et al. in \cite{horowitz1975non}. FORE provides the advantage of filter frequency placement unlike CI and has been used for narrowband phase compensation in \cite{li2005nonlinear}. The matrices of FORE for Eqn. \ref{eq:reset} where the base linear filter has corner frequency $\omega_r$ are
\begin{equation*}
A_r = -\omega_r,\ B_r = \omega_r,\ C_r = 1,\ D_r = 0,\ A_\rho = 0
\end{equation*}

FORE was generalized in \cite{guo2009frequency} to obtain Generalized FORE (GFORE) which provides the additional freedom of having a non-zero resetting parameter $A_\rho$ and hence controlling the level of reset. This is achieved by using an additional reset parameter $\gamma$ such that $A_\rho = \gamma$, where $\gamma = 1$ results in a linear filter. $\gamma$ is used to influence the amount of nonlinearity and hence phase lag. The influence of $\gamma$ on phase lag and other properties is studied in the next section.

\subsubsection{Second Order Reset Element - SORE}

SORE has been recently developed by Hazelgar et. al. \cite{hazeleger2016second} opening new possibilities for reset controllers in the shape of notch and second order low pass filters. SORE has the advantage of an additional parameter, damping coefficient $\beta_r$ as seen in the base matrix definitions below. This provides an extra degree of freedom in the design of nonlinear resetting element. 2 identical FOREs in series is a special case of SORE with $\beta_r = 1$. The additional parameters $\beta_r$ allows for achieving properties not possible by combination of FOREs. The matrices of SORE as applicable to Eqn. \ref{eq:reset} are given as

\begin{equation*}
A_r=\begin{bmatrix}
0 				& 		1\\
-\omega_r^2 	&     -2\beta_r\omega_r
\end{bmatrix}
,
B_r=\begin{bmatrix}
0\\
\omega_r^2
\end{bmatrix}
\end{equation*}
\begin{equation*}
C_r=\begin{bmatrix}
1 & 0\\
\end{bmatrix},
D_r=\begin{bmatrix}
0
\end{bmatrix}
\end{equation*}
where, 
$\omega_r$ is the corner frequency of the filter;
$\beta_r$ is the  damping coefficient

\section{Frequency domain behaviour of reset elements}
\label{Freqbehav}

\subsection{Generalized SORE (GSORE) and generalization of reset controller}

Hazelgar et. al. introduced SORE in \cite{hazeleger2016second} where both states are reset to zero when the reset condition is met. This can be considered as $traditional\ reset\ control$ system. Such a system similar to FORE provides less flexibility of design and overall design becomes dependent on the base linear system. Hence, we present generalized SORE (GSORE) where $A_{\rho} \in \mathbb{R}^{2\times 2}$ can be an arbitrary resetting matrix. While such a system provides greater freedom in design, 4 additional parameters also add to the complexity of analysis during design. Hence we limit this freedom to one parameter by defining $A_\rho$ similar to the manner in GFORE as
\begin{equation*}
A_\rho = \gamma I_{2\times2}
\end{equation*}

Broadly, reset controllers can be generalized such that resetting matrix $A_\rho$ in Eqn. \ref{eq:reset} is no longer a zero matrix as originally conceived, but is of form

\begin{equation*}
A_\rho = \begin{bmatrix}
\gamma_1 & 0 & . & . & . & . & 0 & 0\\
0 & \gamma_2 & . & . & . & . & 0 & 0\\
0 & 0  & . & . & . & . & \gamma_{nr} & 0\\
0 & 0  & . & . & . & . & 0 & I_{n_{nr}\times n_{nr}}
\end{bmatrix}
\end{equation*}
where $n_r$ and $n_{nr}$ are number of resetting and non-resetting states of overall controller respectively. Each resetting state has its own factor $\gamma$ determining its after reset value as a fraction of its pre-reset value. It must be noted that while this generalized form provides a large degree of freedom in design, this might not be useful or convenient in all cases. This is specially true with loop shaping technique which is generally carried out by experienced engineers and not algorithms; and hence having too many variables for tuning might impede design rather than aid it.

\subsection{Analysis of reset elements using describing function}

Frequency domain behaviour analysis of reset elements in literature has mainly focussed on phase lag reduction. However, loop shaping requires a more comprehensive understanding of the behaviour. This knowledge is also essential for design of CgLp presented in the next section. This analysis is carried out using describing function method explained in Sec \ref{resetprops} for the reset elements. Describing function based frequency behaviour is obtained for GSORE for different values of $\gamma \in [-1,1]$ and is shown in Fig. \ref{fig:phasevalred}. There are three important characteristics, two in gain and one in phase, which needs to be noted. While the change in phase behaviour has been studied greatly in literature, the effect of reset on gain is not found in literature to the best of authors' knowledge.

\begin{figure}
	\centering
	\includegraphics[trim = {2cm 0 2.5cm 0}, width=1\linewidth]{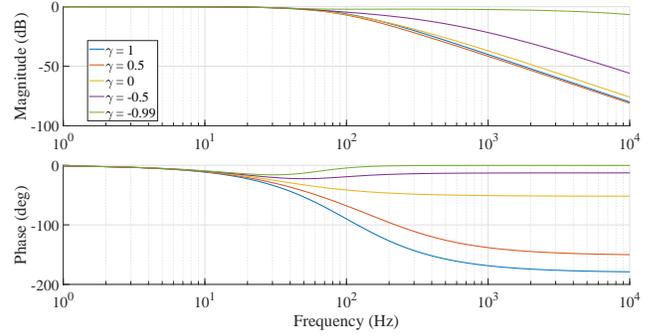}
	\caption{Describing function based frequency response of GSORE for different values of $\gamma$ with $\omega_r = 2\pi100$ and $\beta_r = 1$}
	\label{fig:phasevalred}	
\end{figure}

\begin{itemize}
	\item 
	Shift in corner frequency: 	From the figure, it can be seen that for values of $\gamma \in [0,1]$, gain behaviour of reset element is similar to that of its linear counterpart ($\gamma = 1$). However, for values of $\gamma < 0$, while the slope of gain is still $-40\ db/decade$ at high frequencies, there is a shift in the corner frequency of the filter. While this is shown here for GSORE, this is also true for filters of other orders \cite{saikumargeneralized}. This shift is parametrized as fraction $\alpha$ where 
	$$
	\alpha = \dfrac{\textnormal{Corner frequency of reset element}}{\textnormal{Corner frequency of base linear element}}
	$$
	The value of $\alpha$ as a function of $\gamma$ is plotted in Fig. \ref{fig:cornerfrac} for GFORE and GSORE ($\beta_r = 1$). While the value of $\alpha$ can be used to modify the base linear system to ensure that the corner frequency of reset filter is at the desired value, these $\alpha$ values are close to 1 for values of $\gamma \geq 0$. However, they increase in an almost exponential manner as value of $\gamma$ is reduced further. This limits the values of $\gamma$ for which GFORE and GSORE can be effectively used in practice.
	\item 
	Phase lag reduction: While the difference in gain is only seen for lower values of $\gamma$, reduction in phase lag is seen to be sensitive and is seen for all values of $\gamma < 1$. The phase lag of GSORE for different values of $\gamma$ is shown in Fig. \ref{fig:phvsal}. Phase lag achieved with GFORE is also shown in the same figure for comparison. It can be seen that large phase lag reductions are seen for smaller values of $\gamma$ with phase lag being zero at $\gamma = -1$. However, due to the corresponding change in corner frequency of GSORE as seen in Fig. \ref{fig:phasevalred}, use of these generalized elements becomes limited.
	\item 
	Change in damping factor: Another interesting characteristic of GSORE is seen in change in the damping factor of the designed linear filter and achieved GSORE filter. This is shown in Fig. \ref{fig:QF} for different values of $\beta_r$, where it can be seen that even for $\beta_r = 0$ (resulting in a $Q\ factor = \infty$ in the case of linear filter), the resonance peak is less than $10\ dB$. Although change in gain plot vs $\beta_r$ is negligible, there is change in phase plot with changing $\beta_r$ values and this can be used advantageously to obtain a sharp change in phase without the cost of a resonance peak. This additional advantage is only seen with GSORE due to presence of damping factor $\beta_r$ and not in GFORE.
\end{itemize}

\begin{figure}
	\centering
	\includegraphics[trim = {3.5cm 0 3.5cm 0}, width=1\linewidth]{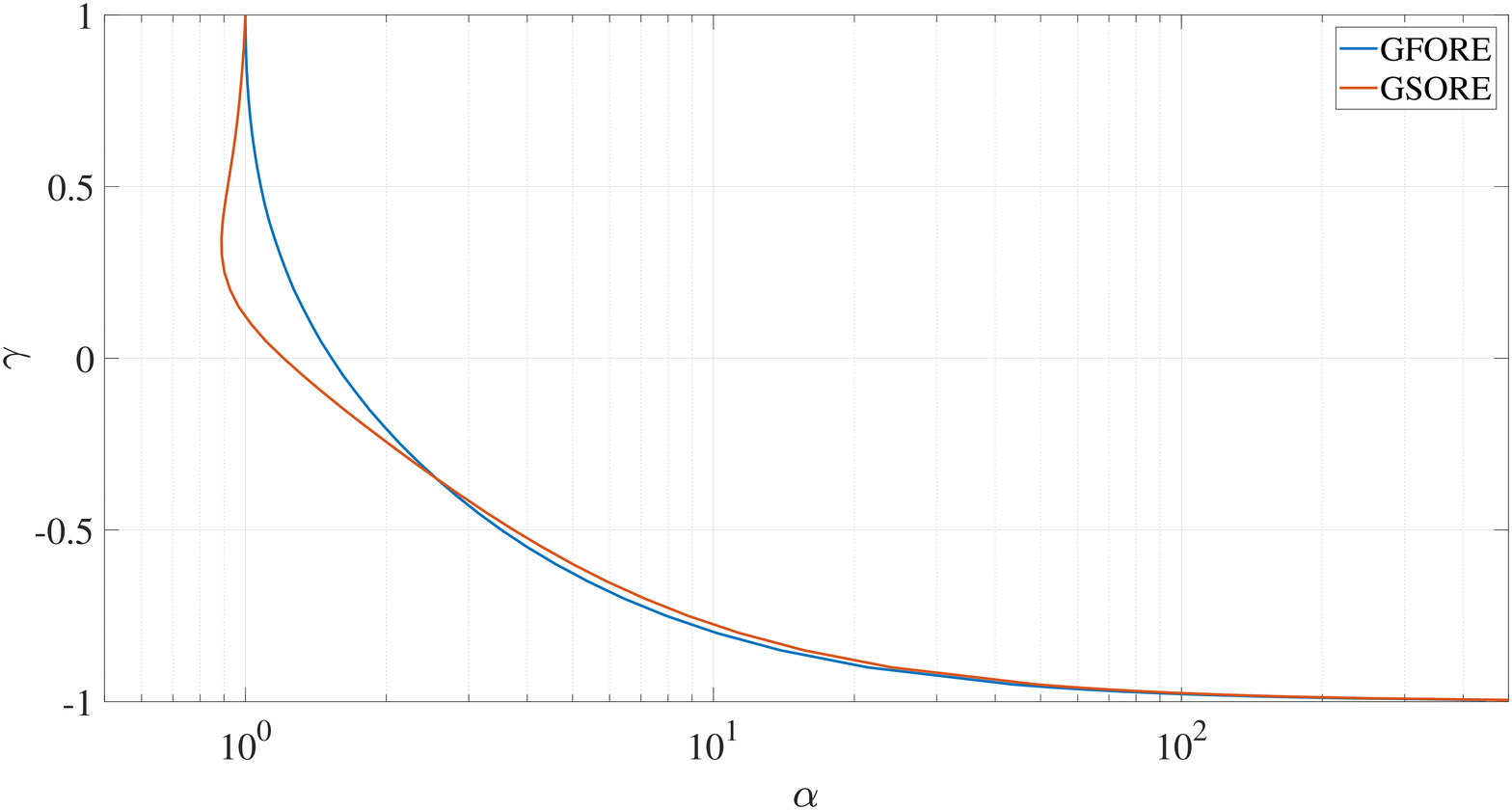}
	\caption{Fraction $\alpha$ denoting extent of change in corner frequency as a function of $\gamma$ for GFORE and GSORE ($\beta_r = 1$)}
	\label{fig:cornerfrac}	
\end{figure}

\begin{figure}
	\centering
	\includegraphics[trim = {3cm 0 3.5cm 0}, width=1\linewidth]{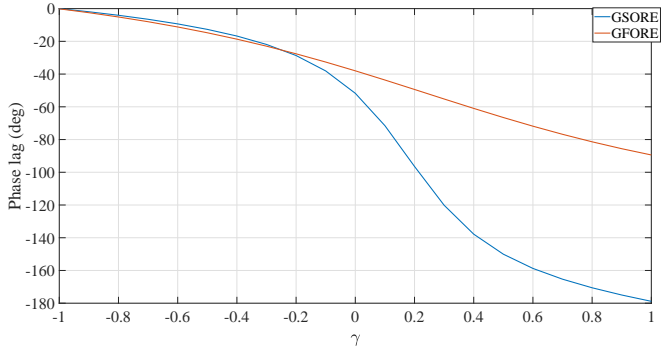}
	\caption{Reduction in phase lag with reset for both GSORE and GFORE}
	\label{fig:phvsal}	
\end{figure}

\begin{figure}
	\centering
	\includegraphics[trim = {3cm 0 2.5cm 0}, width=1\linewidth]{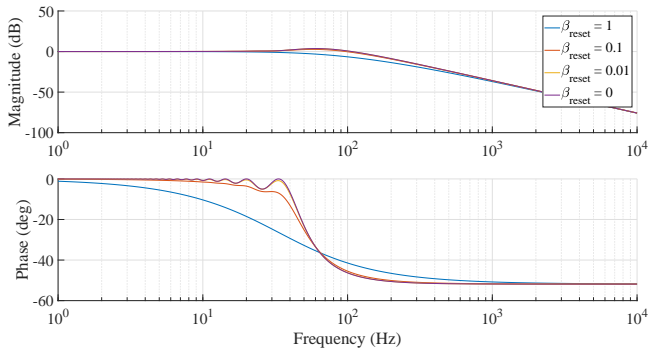}
	\caption{Change in damping value of GSORE with $\gamma = 0$}
	\label{fig:QF}	
\end{figure}

In the above frequency domain analysis of reset elements, sinusoidal input describing function analysis has been used to obtain the frequency response. While this pseudo-linear technique is useful, it is only an approximation technique. To verify the accuracy of this method, we obtained the frequency response of GSORE directly by applying chirp and step inputs and using the $tfestimate$ function of MATLAB and comparing the response to the one obtained from describing function. The coherence $C_{xy}$ which gives a measure of the accuracy of obtained frequency response using $tfestimate$ is also obtained and plotted in Fig. \ref{fig:simul1}. The plots show good match between the describing function based results and those obtained through estimation in MATLAB.

\begin{figure}
	\centering
	\includegraphics[width=1\linewidth]{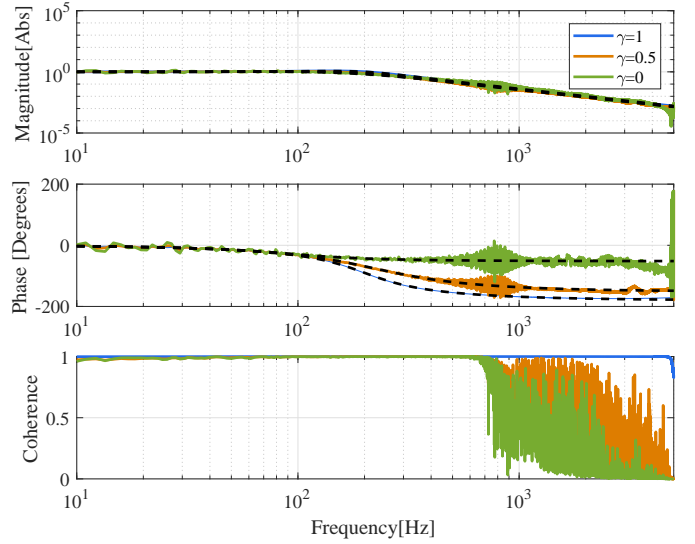}
	\caption{Gain, Phase and Coherence relation for different values of $\gamma$. Dashed line represents the values from the describing function. $\omega_r = 2\pi200$}
	\label{fig:simul1}	
\end{figure}

\end{theorem}

\section{Constant-gain Lead-phase (CgLp)}
\label{CgLpsec}

Reset is used in controls for its phase lag reduction. SORE helps in this regard, with generalization further providing the freedom to choose the level of reset and hence the level of nonlinearity introduced. However, low pass filters are generally used in controls at high frequency for noise attenuation. While these can be replaced by reset low pass filters in the form of GFORE or GSORE advantageously, this use of reset results only in phase lag reduction. Reset for phase lead which can be used advantageously in region of bandwidth has not been explored sufficiently in literature. The main works in this regard as noted earlier are \cite{li2005nonlinear} and \cite{li2011reset}. Here, we introduce a new reset element termed Constant in gain Lead in phase (CgLp) which uses GFORE (or GSORE) to provide broadband phase compensation in the required range of frequencies.

\subsection{Definition}

Broadband phase compensation is achieved in CgLp by using a reset lag filter $R$ (GFORE or GSORE) in series with a corresponding order linear lead filter $L$ as given below.

\begin{equation}
R(s) = \dfrac{1}{\cancelto{\gamma}{(s/\omega_{r\alpha})^2 + (2s\beta_r/\omega_{r\alpha}) + 1}} \ \ \ \ \ \ or\  \dfrac{1}{\cancelto{\gamma}{s/\omega_{r\alpha} + 1}}
\label{Reset_lag}
\end{equation}
and 
\begin{equation}
L(s) = \dfrac{(s/\omega_r)^2 + (2s\beta_r/\omega_r) + 1}{(s/\omega_f)^2 + (2s/\omega_f) + 1} \ \ \ \ \ \ or\ \dfrac{s/\omega_r + 1}{s/\omega_f + 1}
\end{equation}
correspondingly with $\omega_f >> \omega_r, \omega_{r\alpha}$. The arrow indicates the resetting nature of $R$. $\omega_{r\alpha} = \omega_r/\alpha$ accounting for the shift in corner frequency with reset as noted in Sec. \ref{resetprops} and can be obtained from Fig. \ref{fig:cornerfrac} for the chosen value of $\gamma$.

The reset state matrices of CgLp using GFORE are given below as 

\[
A_r=\begin{bmatrix}
-\omega_{r\alpha} &	0\\
\omega_f & -\omega_f
\end{bmatrix}
,
B_r=\begin{bmatrix}
\omega_{r\alpha}\\
0
\end{bmatrix}
\]
\[
C_r=\begin{bmatrix}
\dfrac{\omega_f}{\omega_r} & \Bigg(1 - \dfrac{\omega_f}{\omega_r}\Bigg)
\end{bmatrix},
D_r=\begin{bmatrix}
0
\end{bmatrix}\]
\[
A_\rho = \begin{bmatrix}
\gamma & 0\\
0 & 1
\end{bmatrix}
\]

and CgLp using GSORE are given below as

\[
A_r=\begin{bmatrix}
0 				& 		1	&	0	&	0\\
-\omega_{r\alpha}^2 	&     -2\beta_r\omega_{r\alpha}	&	0	&	0\\
0	&	0	&	0	&	1\\
1	&	0	&	-{\omega_f}^2	&	-2\omega_f
\end{bmatrix}
,
B_r=\begin{bmatrix}
0\\
\omega_{r\alpha}^2\\
0\\
0
\end{bmatrix}
\]
\[
C_r=\begin{bmatrix}
\dfrac{{\omega_f}^2}{{\omega_r}^2}	&	0	&	\Bigg({\omega_f}^2 - \dfrac{{\omega_f}^4}{{\omega_r}^2}\Bigg)	&	\Bigg(\dfrac{2\beta_r{\omega_f}^2}{\omega_r} - \dfrac{2{\omega_f}^3}{{\omega_r}^2}\Bigg)\\
\end{bmatrix}
\]
\[
D_r=\begin{bmatrix}
0
\end{bmatrix},
A_\rho = \begin{bmatrix}
\gamma I & 0\\
0 & I
\end{bmatrix}
\]

These matrices are used to describe CgLp element in the general form of reset as in Eqn. \ref{eq:reset}.

\begin{figure}
	\centering
	\includegraphics[trim = {3.5cm 0 3.5cm 0}, width=1\linewidth]{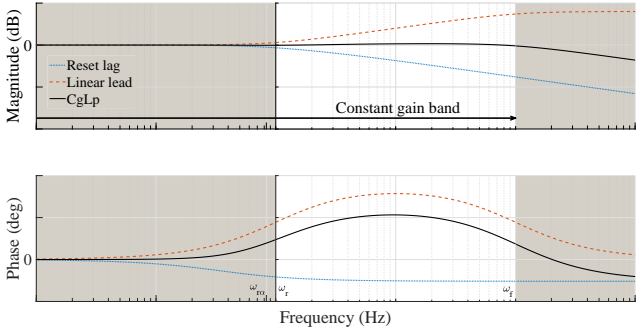}
	\caption{Broadband phase lead achieved with CgLp using GSORE in range $[\omega_r,\omega_f = 100\omega_r]$ with $\gamma = 0$ and $\beta_r = 1$. For $\gamma = 0$, value of $\alpha = 1.2$ is obtained from Fig. \ref{fig:cornerfrac} resulting in $\omega_{r\alpha} = 0.8333\omega_r$}
	\label{fig:CgLp}	
\end{figure}

In the conventional case, with both lag and lead filters placed at the same frequency we have $$\textnormal{Gain of linear lead filter} - \textnormal{Gain of linear lag filter} = 0\ dB$$ $$\textnormal{Phase of linear lead filter} - \textnormal{Phase of linear lag filter} = 0^\circ$$
However, with reset applied to lag filter $R$, assuming that we account for change in corner frequency with correct value of $\alpha$, the gains still cancel each other. But with phase,
$$
\textnormal{Phase of linear lead filter} - \textnormal{Phase of reset lag filter} > 0^\circ
$$
resulting in phase lead in range $[\omega_r,\omega_f]$ where the value of phase lead obtained is dependent on choice of GFORE or GSORE for design and also on value of $\gamma$ providing freedom of choice to the control engineer. Broadband phase lead achieved through CgLp is shown in the frequency response of an example CgLp element in Fig. \ref{fig:CgLp}.

\subsection{Comparison between CgLp using GSORE and GFORE}

\begin{figure}
	\centering
	\includegraphics[trim = {2.5cm 0 3.5cm 0}, width=1\linewidth]{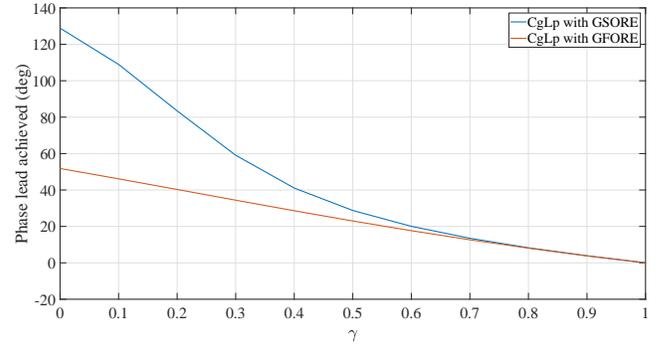}
	\caption{Phase lead obtained through CgLp for different values of $\gamma$}
	\label{fig:gforevsgsore}	
\end{figure}

A first order lead filter can provide maximum of $90^\circ$ phase lead and a corresponding reset lag filter $GFORE$ can have a phase lag of $0^\circ$ at $\gamma = -1$ as seen in Fig. \ref{fig:phvsal}, resulting in a maximum phase compensation of $90^\circ$. Similarly, with a second order lead filter and GSORE, a phase lead of up to $180^\circ$ can be achieved. However, as noted earlier, for values of $\gamma < 0$, the value of $\alpha$ in Fig. \ref{fig:cornerfrac} increases exponentially and becomes very large. Hence we limit the value of $\gamma \geq 0$ in this study. This correspondingly limits the maximum phase compensation that can be achieved to $51.9^\circ$ and $128.1^\circ$ for CgLp with GFORE and GSORE respectively. The phase lead achieved at different values of $\gamma$ is shown in Fig. \ref{fig:gforevsgsore}.

Resetting parameter $\gamma$ controls the deviation from linear configuration and is associated with the level of nonlinearity. Although we have considered describing function to analyse frequency domain behaviour, the resetting action results in higher order harmonics. And we can assume that higher the level of nonlinearity, higher is the presence and negative effect of harmonics on system performance. This leads to a trade-off between phase compensation and effects of harmonics. So it can be said that to minimize the effect of harmonics, GSORE is the obvious choice in CgLp, since it minimizes $|1 - \gamma|$ for the same phase lead compensation and reduces the inherent trade-off. However, with GSORE, two states are being reset and this could lead to more unwanted harmonics compared to GFORE. These theories need to be investigated further in future work to determine optimal design of CgLp.

\subsection{CgLp in PID framework}

The phase lead achieved over a large range of frequencies with a corresponding $0\ dB/dec$ slope line in gain allows CgLp to overcome Bode's gain phase relation which limits linear controllers. While this broadband phase compensation of CgLp can be used advantageously in different areas of control, here we deal with integrating CgLp into framework of PID to improve performance in terms of precision and tracking for high-tech applications.

The general structure of series PID as used in the industry for loop shaping is given as:
\begin{equation}
\label{eqPID}
PID = K_p\Bigg(\frac{s + \omega_i}{s}\Bigg)\Bigg(\frac{1 + \frac{s}{\omega_d}}{1 + \frac{s}{\omega_t}}\Bigg)\Bigg(\frac{1}{1 + \frac{s}{\omega_f}}\Bigg)
\end{equation}
where $\omega_i$ is the frequency at which integrator action is terminated, $\omega_d$ and $\omega_t$ are the starting and  taming frequencies of differentiator action, and $\omega_f$ is corner frequency of low pass filter used to attenuate noise at high frequencies with $\omega_i < \omega_d < \omega_t < \omega_f$. Additional notch, anti-notch or low pass filters may also be included in design of PID depending on system to be controlled. In this structure of PID, it is easy to interpret the different functions of its parts. While the integrator creates high gain at low frequencies improving tracking, low pass filters reduces gain at high frequencies improving noise attenuation and hence precision. Derivative action adds phase in region of bandwidth and hence contributes to stability and robustness. 

To ensure that maximum phase achieved through derivative action coincides with bandwidth ($\omega_c$), $\omega_d$ and $\omega_t$ are chosen as $\omega_d = \omega_c/a$ and $\omega_t = a\omega_c$ where $a > 1$. The value of scale $a$ determines the phase provided by derivative. \cite{schmidt2014design} suggests $a = 3$ as rule of thumb for design of PID for high-precision mechatronic systems along with $\omega_i = \omega_c/10$ and $\omega_f = 10\omega_c$.

The value of scale $a$ determines not only the phase margin (PM) and hence stability/robustness of closed-loop system but also the tracking and precision performances. This is because derivative action adds gain at high frequencies and reduces gain at low frequencies hence negatively affecting noise attenuation and tracking respectively. Hence increasing value of $a$ to increase PM negatively affects the other performance aspects and vice-versa. Bode's gain-phase relation which determines this behaviour is a limitation which affects all linear controllers. This fundamental problem can be overcome by integrating CgLp into framework of PID so that either a fraction or complete phase to be added at bandwidth can be obtained through CgLp.

Hence, 3 different extreme scenarios can be considered where some of the performance criteria can be improved while ensuring that the others are not compromised in the process. These are listed out below.

\begin{itemize}
	\item 
	The values of $\omega_d$ and $\omega_t$ can be fixed to the values obtained for linear PID and CgLp designed to add required additional phase and hence improve stability and robustness without affecting precision, tracking or bandwidth.
	\item 
	CgLp can be designed first to provide part of the phase resulting in a smaller scale $a$ for derivative action (to obtain same PM) which should result in improved tracking and precision without affecting stability and bandwidth.
	\item 
	CgLp can be designed to provide part of the phase again as in the second case, but instead of improving precision, the closed loop bandwidth of the system can be increased which thereby increases tracking as well without affecting stability or precision.
\end{itemize}

\begin{figure}
	\centering
	\includegraphics[width=1\linewidth]{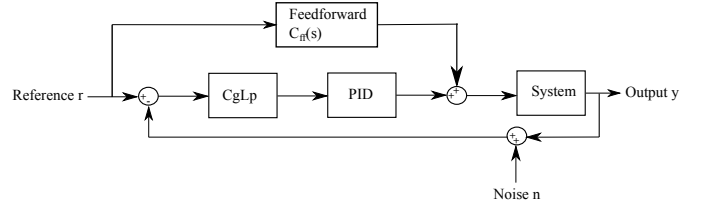}
	\caption{General block diagram of CgLp-PID controlled closed loop system}
	\label{fig:Block}	
\end{figure}

While these are the extreme cases portrayed, intermediate options where stability, robustness, tracking, precision and bandwidth are simultaneously improved can also be explored. Only the second option (same stability and bandwidth, improved tracking and precision performance) and third option (same stability and precision, improved bandwidth and tracking performance) are explored further in this paper and the rest will have to be part of future work. The general block-diagram of CgLp-PID can be visualized as shown in Fig. \ref{fig:Block} which also consists of a feedforward block for improved tracking performance.

\section{Application on precision mechatronic system}
\label{Application}

\subsection{Design of CgLp-PID for performance comparison}

The world of precision high-tech industry is pushing towards faster, more precise and better tracking systems constantly. Hence we explore the option of using CgLp-PID to improve these performance aspects while maintaining stability margins at same level. Since the controllers are designed for performance comparison, we have established a baseline in design. For design of PID part, $\omega_i = \omega_c/10$ and $\omega_f = 10\omega_c$ which are rules of thumb are used for all controllers designed. For design of CgLp, we have chosen $\omega_r = \omega_c$ as common. CgLp designs using both GFORE and GSORE are considered for improved performance in terms of tracking and precision improvement, while only CgLp using GFORE is considered and tested for improvement in bandwidth and tracking.

\subsubsection{Tracking and precision improvement}
\label{trackprecise}

In this part, we look at CgLp-PID design for improvement in tracking and precision without compromising stability and bandwidth. These are the steps followed in detail.

\begin{enumerate}[(i)]
	\item For CgLp element, choose $\omega_r = \omega_c$, $\omega_f = 10\omega_c$.
	\item Choose value of $\gamma \in [0,1]$ and correspondingly choose value of $\alpha$ from Fig. \ref{fig:cornerfrac} to calculate $\omega_{r\alpha}$ to account for shifting of corner frequency.
	\item Compute phase added by CgLp element at $\omega_c$ using describing function analysis as $Ph_{nl}$.
	\item For PID, choose $\omega_i = \omega_c/10$, $\omega_f = 10\omega_c$.
	\item Additional phase that needs to be obtained through derivative action within PID = $Required\ PM - Ph_{nl}$.
	\item Choose scale $a$ and obtain values of $\omega_d$ and $\omega_t$ such that this additional phase is achieved to ensure overall $Required\ PM$.
\end{enumerate}

\color{black}

\subsubsection{Bandwidth and tracking improvement}
\label{bandprecise}

Here, closed-loop bandwidth of system is increased using CgLp-PID design which results in improvement in tracking. This improvement is achieved without compromising precision and stability. While the steps in the previous case are straightforward, this is not true here with multiple iterations required. These are the steps followed to design a separate set of controllers. Since the controllers are designed to achieve same precision, it is necessary to calculate the precision achieved with linear PID. This can be obtained from the open-loop gain value at a sufficiently high frequency $\omega_{high}$, where the behaviour is asymptotic. Let this value be $G_{pre}$.

\begin{enumerate}[(a)]
	\item Choose bandwidth $\omega_c$.
	\item Follow steps (i) - (vi) from Sec. \ref{trackprecise}.
	\item Calculate gain of open-loop at $\omega_{high}$ to check if same precision is achieved (with a margin of error). If achieved precision is higher than $G_{pre}$, increase $\omega_c$ and go back to step (b). If lower, reduce $\omega_c$ and go back to step (b). 
\end{enumerate}

\color{black}

\subsection{Precision positioning stage}

\begin{figure}
	\centering
	\includegraphics[width=1\linewidth]{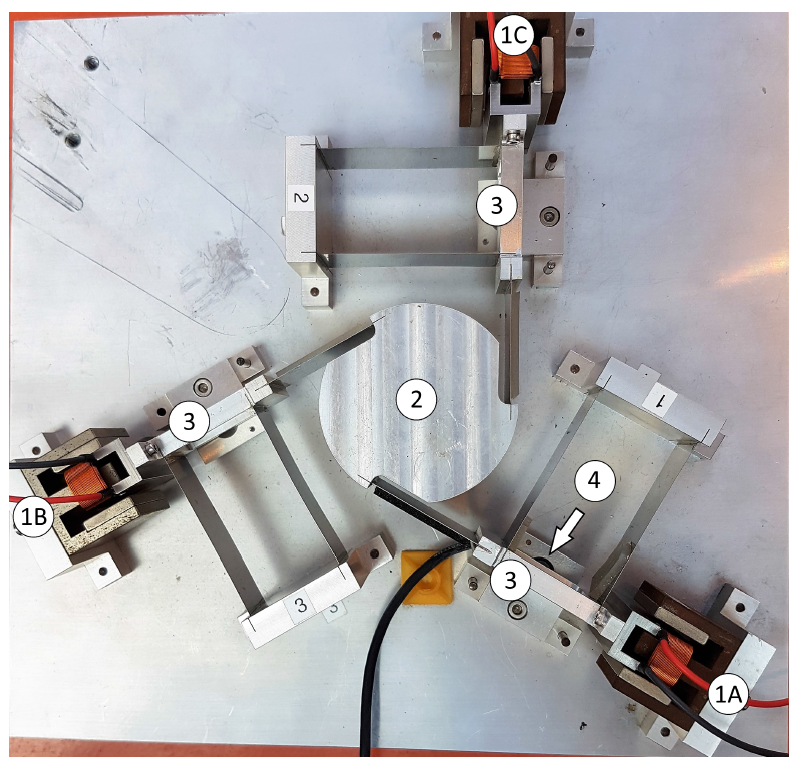}
	\caption{3 DOF planar precision positioning `Spyder' stage. Voice coil actuators 1A, 1B and 1C control 3 masses (indicated as 3) which are constrained by leaf flexures. The 3 masses are connected to central mass (indicated by 2) through leaf flexures. Linear encoders (indicated by 4) placed under masses '3' provide position feedback.}
	\label{fig:setup}	
\end{figure}

A precision planar positioning stage shown in Fig. \ref{fig:setup} is used for validation and performance analysis of developed CgLp-PID controllers. 2 sets of controllers are designed for both the cases considered above in terms of performance improvement. For sake of simplicity, only one of the actuators (1A) is considered and used for controlling position of mass '3' attached to same actuator resulting in a SISO system. All designed controllers are implemented on FPGA of NI Myrio system to achieve fast real-time control. LM388 linear power amplifier is used to power the actuator and Mercury M2000 linear encoder is used to obtain position feedback with a resolution of $100\ nm$.

\begin{figure}
	\centering
	\includegraphics[trim = {2.5cm 0 3.5cm 0}, width=1\linewidth]{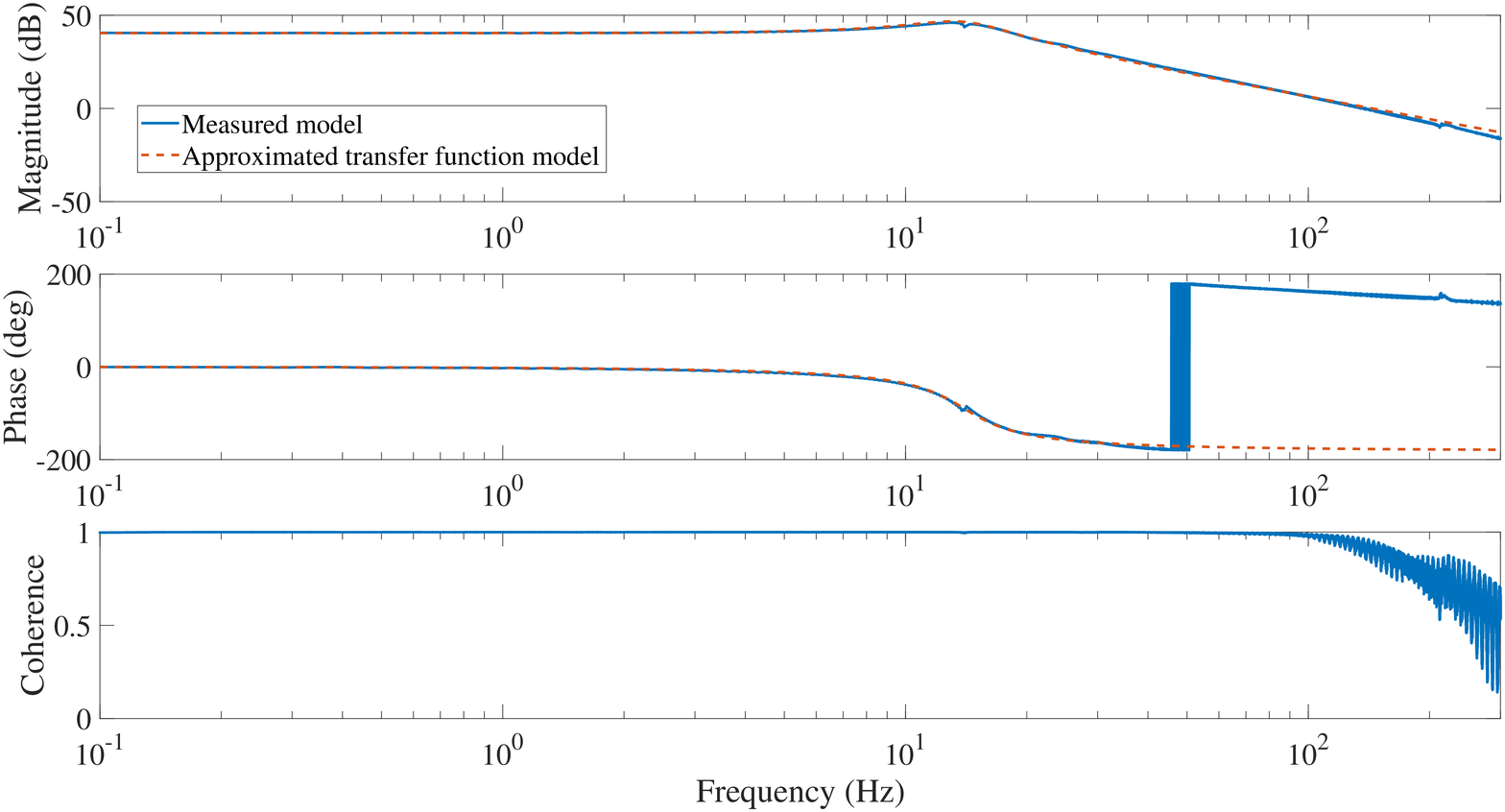}
	\caption{Frequency response data of system as seen from actuator '1A' to position of mass '3' attached to same actuator. Frequency response of simplified estimated transfer function is also plotted for comparison.}
	\label{fig:frf}	
\end{figure}

In keeping with industry techniques, frequency response data of system is obtained by applying a chirp signal and this is shown in Fig. \ref{fig:frf}. The system behaviour is similar to that of a collocated double mass-spring-damper system with additional dynamics at higher frequencies. The system can, however, be simplified to a second-order system as given below. The frequency response of this simplified second order system is also given in Fig. \ref{fig:frf} for comparison. This transfer function is used for stability analysis using theorems presented in Sec. \ref{resetstab}.

\begin{equation}
\label{eq:system}
System\ G(s) = \frac{1.429e8}{175.9s^2 + 7738s + 1.361e6}
\end{equation}

Although we design the controllers using frequency response data with the assumption that the system is linear, it must be noted that the spring stiffness of the leaf flexures used is not constant over the full stroke length and hence results in variations in gain and resonance frequencies over the tested stroke.

\subsection{Designed controllers}

Controllers are to be designed to achieve a bandwidth of $\omega_c = 100\ Hz$ along with PM of $30^\circ$. The phase of system at $\omega_c$ is $-195^\circ$ and hence a phase lead of $45^\circ$ needs to be achieved by all the designed controllers.

\color{black}

\subsubsection{Controllers using Reset Integrator}

Reset integrator has been popularly used in literature for its phase lag reduction advantage. As a benchmark for comparison, 6 controllers are designed for different values of $\gamma$ with the integrator part of Eqn. \ref{eqPID}, i.e., $(1/s)$ being reset and the rest of the equation used as the linear controller. The controllers are designed using the same 6 steps mentioned for design of CgLp-PID with the modification that the phase compensation comes from the reduced phase lag of the resetting integrator and not from CgLp. The value of scale $a$ obtained in step (vi) for each corresponding value of $\gamma$ is provided in Table. \ref{Tabagamma}.

\color{black}

\subsubsection{Controllers for increase in tracking and precision}

6 controllers are designed with CgLp using GFORE and 6 other controllers with CgLp using GSORE according to the steps mentioned in Sec. \ref{trackprecise} for values of $\gamma$ ranging from 1 resulting in purely linear controller to 0 in steps of 0.2. The value of scale $a$ obtained in step (vi) for each corresponding value of $\gamma$ is provided in Table. \ref{Tabagamma}. 

\begin{table}
	\centering
	\begin{tabular}{|c|c|c|c|}
		\hline
		 & Reset Integrator & CgLp with GFORE & CgLp with GSORE\\ \hline\hline
		$\gamma$ & \multicolumn{3}{|c|}{$scale\ \ a$}\\ \hline\hline
		1.0 & 2.9 & 2.9 & 2.9 \\ \hline
		0.8 & 2.35 & 2.63 & 2.27 \\ \hline
		0.6 & 1.89 & 2.43 & 1.81 \\ \hline
		0.4 & 1.52 & 2.27 & 1.46 \\ \hline
		0.2 & 1.23 & 2.12 & 1.24 \\ \hline
		0.0 & 1.01 & 1.98 & 1.09 \\ \hline
	\end{tabular}
	\caption{Values of $scale\ a$ used in derivative action corresponding to value of $\gamma$ used in the designed controllers. In combination, they achieve phase lead of $45^\circ$ at $\omega_c = 100\ Hz$.}
	\label{Tabagamma}
\end{table}

\begin{figure}
	\centering
	\includegraphics[trim = {3cm 0 3.5cm 0}, width=1\linewidth]{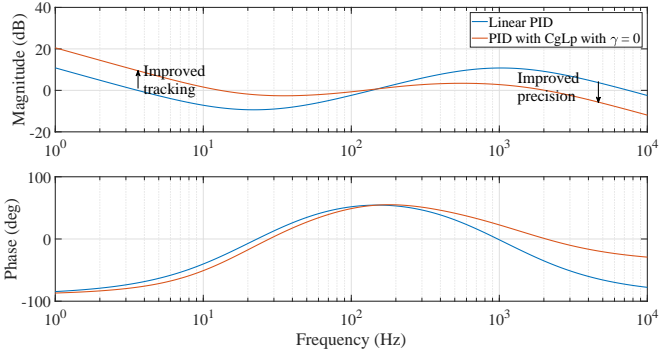}
	\caption{Frequency response of 2 controllers designed with GFORE for improvement in precision and tracking obtained through describing function analysis.}
	\label{fig:contro}	
\end{figure}

The frequency responses of designed controllers for the two extreme cases of CgLp with GFORE, i.e., $\gamma = 1$ (resulting in linear PID) and $\gamma = 0$ obtained through describing function are shown in Fig. \ref{fig:contro}. It can be seen that for $\gamma = 0$, since a large phase lead is achieved through CgLp element, phase lead required from derivative action is less resulting in a much smaller value of scale $a$ as seen in Table. \ref{Tabagamma}. Further, this value of scale $a$ is even more reduced in the case of CgLp with GSORE and also reset integrator. This results in reduced gain at high frequencies and increased gain at low frequencies and hence better precision and tracking respectively are to be expected. 

\color{black}

\subsubsection{Controllers for increase in bandwidth and tracking}

\begin{figure}
	\centering
	\includegraphics[trim = {2cm 0 3.5cm 0}, width=1\linewidth]{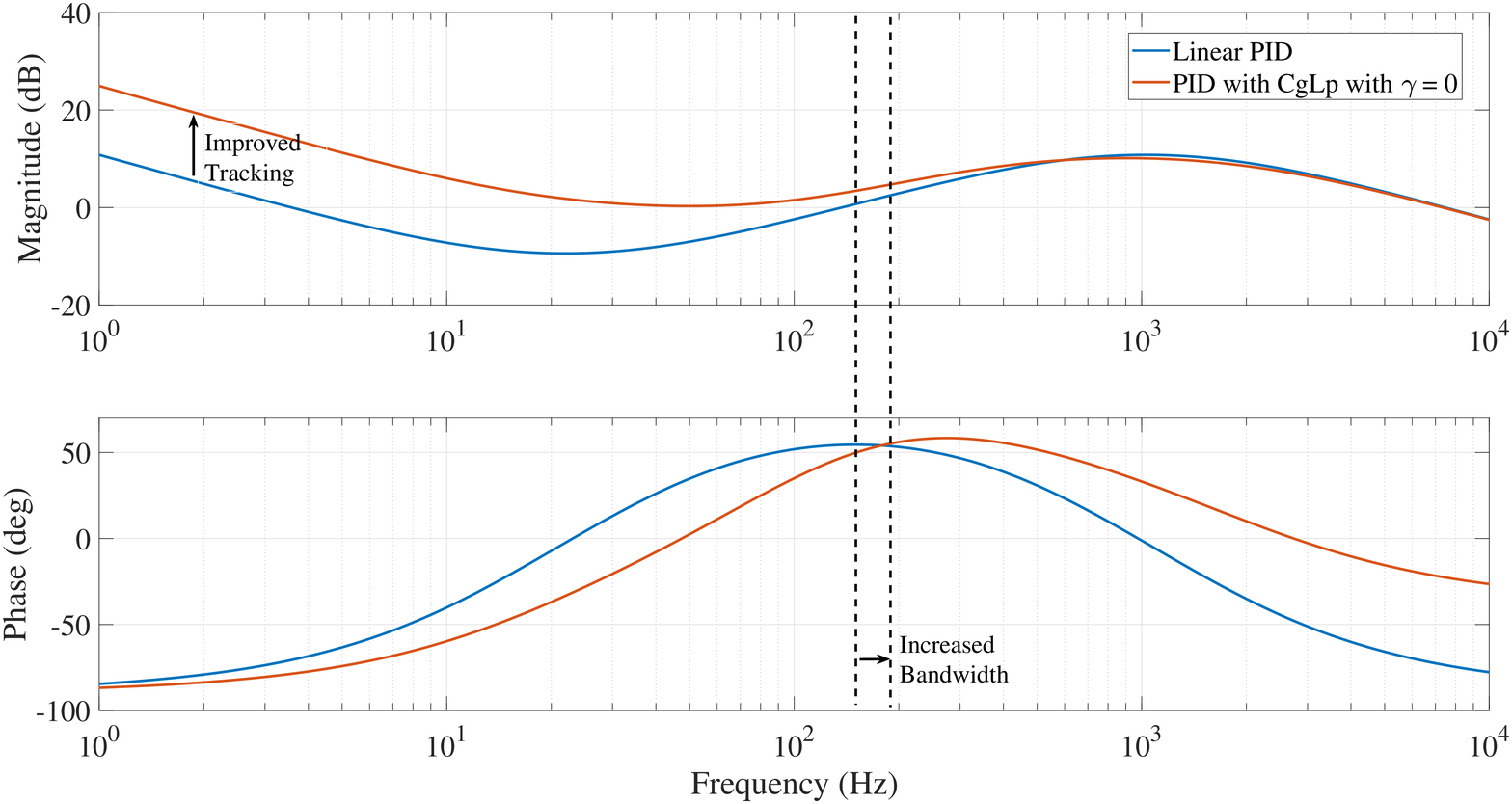}
	\caption{Frequency response of 2 controllers designed for improved bandwidth and tracking obtained through describing function analysis.}
	\label{fig:contro2}	
\end{figure}

A separate set of controllers are designed to obtain the same precision. In this case, only controllers with CgLp using GFORE are designed for comparison. For this, the controller with $\gamma = 1$ which results in linear PID is designed at $\omega_c = 100\ Hz$ as before. The theoretical precision that can be achieved with this is estimated by obtaining the open-loop gain at $10\ KHz$ and this is found to be $G_{pre} = -76.18\ dB$. This along with $PM = 30^\circ$ is used as reference to design 5 other controllers for values of $\gamma$ from $0.8$ to $0$ in intervals of $0.2$ by following the steps given in Sec. \ref{bandprecise}. The bandwidth and scale $a$ values obtained for the designed controllers are provided in Table. \ref{Tabagamma2}. The frequency response of the two extreme cases is obtained using describing function analysis and shown in Fig. \ref{fig:contro2}. The additional increase in tracking as compared to the extreme cases of Fig. \ref{fig:contro} can be seen due to the increase in bandwidth. Further, it can also be noticed that since phase of the system decreases at higher frequencies, additional phase has to be generated to ensure required $PM$. This is the reason that values of scale $a$ in Table. \ref{Tabagamma} for CgLp with GFORE and Table. \ref{Tabagamma2} do not match each other.

\begin{table}
	\centering
	\begin{tabular}{|c|c|c|}
		\hline
		$\gamma$ & Bandwidth (Hz) & $scale\ a$\\ \hline\hline
		1.0 & 100 & 2.9 \\ \hline
		0.8 & 107 & 2.73 \\ \hline
		0.6 & 113 & 2.60 \\ \hline
		0.4 & 118.5 & 2.47 \\ \hline
		0.2 & 123 & 2.35 \\ \hline
		0.0 & 127 & 2.27 \\ \hline
	\end{tabular}
	\caption{Values of bandwidth $\omega_c$ and $scale\ a$ used in derivative action corresponding to value of $\gamma$ used in CgLp with GFORE. In combination, they achieve $PM$ of $30^\circ$ and same open-loop gain value at $\omega_{high} = 10\ KHz$ resulting in same precision theoretically.}
	\label{Tabagamma2}
\end{table}

\color{black}

\subsection{Results}

All designed controllers are discretized with sampling frequency of $10\ KHz$ and implemented on the practical setup. Tracking and precision performance aspects are analysed. For the purpose of tracking a fourth order prefiltered trajectory is planned as explained in \cite{lambrechts2005trajectory} for a triangular reference of peak-to-peak amplitude of $1\ mm$. The inverse of estimated system transfer function of Eq. \ref{eq:system} is made strictly proper with a third order filter with corner frequency of $1000\ Hz$ (same corner frequency as that of LPF used in PID) and is used as feedforward controller ($C_{ff}(s)$). The RMS error values for the controllers designed to check improvement in tracking and precision are given in Table. \ref{Tabperformance}, while the results for the second set of controllers designed for evaluating improvement in bandwidth and tracking are provided in Table. \ref{Tabperformance2}. For evaluating precision, although the sensor signal is noisy, additional noise is added at the point shown in Fig. \ref{fig:Block} in the form of uniform gaussian noise of maximum amplitude $5000\ nm$. The reference is made zero for this and the output precision analysed. The RMS and maximum error values obtained are given in the same tables, Table. \ref{Tabperformance} and Table. \ref{Tabperformance2}.

\color{black}

\begin{table*}
	\centering
	\begin{tabular}{|c|c|c|c|c|c|c|c|c|c|}
		\hline
		 & \multicolumn{3}{c|}{Reset Integrator} & \multicolumn{3}{|c|}{CgLp wiht GFORE} & \multicolumn{3}{c|}{CgLp wiht GSORE} \\ \hline \hline
		$\gamma$ & Tracking & \multicolumn{2}{c|}{Precision} & Tracking & \multicolumn{2}{|c|}{Precision} & Tracking & \multicolumn{2}{c|}{Precision} \\ \hline
		 & $e_{RMS}$ & $e_{RMS}$ & $max(|e|)$ & $e_{RMS}$ & $e_{RMS}$ & $max(|e|)$ & $e_{RMS}$ & $e_{RMS}$ & $max(|e|)$ \\ 
		 & $(100\ nm)$ & $(100\ nm)$ & $(100\ nm)$ & $(100\ nm)$ & $(100\ nm)$ & $(100\ nm)$ & $(100\ nm)$ & $(100\ nm)$ & $(100\ nm)$ \\ \hline \hline
		1.0 & 162.9 & 814.1 & 3300 & 162.9 & 814.1 & 3300 & 162.9 & 814.1 & 3300 \\ \hline
		0.8 & 745.9 & 516.7 & 2700 & 149.0 & 569.8 & 2400 & 131.4 & 434.3 & 2000 \\ \hline
		0.6 & 895.7 & 496.6 & 2000 & 140.5 & 522.4 & 2000 & 126.4 & 387.7 & 1700  \\ \hline
		0.4 & 936.5 & 369.3 & 1500 & 141.7 & 486.8 & 1900 & 126.6 & 387.3 & 1500 \\ \hline
		0.2 & 971.7 & 310.0 & 1500 & 143.4 & 491.6 & 1900 & 130.2 & 394.3 & 1700 \\ \hline
		0.0 & 1028.9 & 341.0 & 1500 & 145.9 & 504.3 & 1700 & 138.6 & 408.3 & 1800 \\ \hline
	\end{tabular}
	\caption{Performance indices of controllers designed for improvement in tracking and precision in comparison with value of $\gamma$ defining $CgLp$ element.}
	\label{Tabperformance}
\end{table*}

\begin{table}
	\centering
	\begin{tabular}{|c|c|c|c|}
		\hline
		$\gamma$ & Tracking & \multicolumn{2}{|c|}{Precision} \\ \hline
		& $e_{RMS}$ & $e_{RMS}$ & $max(|e|)$ \\ 
		& $(100\ nm)$ & $(100\ nm)$ & $(100\ nm)$  \\ \hline \hline
		1.0 & 162.9 & 814.1 & 3300 \\ \hline
		0.8 & 144.8 & 591.6 & 3200 \\ \hline
		0.6 & 136.7 & 531.7 & 2900 \\ \hline
		0.4 & 134.6 & 523.8 & 2300 \\ \hline
		0.2 & 137.1 & 529.9 & 2300 \\ \hline
		0.0 & 144.9 & 561.1 & 2800 \\ \hline
	\end{tabular}
	\caption{Performance indices of controllers  designed for improvement in tracking and bandwidth in comparison with value of $\gamma$ defining $CgLp$ element.}
	\label{Tabperformance2}
\end{table}

\color{black}

\color{black}

The performance indices values from Table. \ref{Tabperformance} clearly show the improvement in both tracking and precision as expected with $CgLp$ compared to linear controller ($\gamma = 1$). In fact, in all cases irrespective of value of $\gamma < 1$, tracking and precision have improved significantly. However, it is also seen that while these improvements were also expected for $\gamma = 0$ compared to $\gamma = 0.2$, this is not the case. Performance improves as $\gamma$ is reduced from 1 and is best close to 0.4 for both cases of CgLp controllers and then slightly deteriorates again. The role of the higher order harmonics on performance can also be noted from these results. From Table. \ref{Tabagamma}, it is seen that the value of scale $a$ is the same for CgLp with GFORE and GSORE for $\gamma = 0.4$ and $\gamma = 0.8$ respectively. From describing function analysis, this should result in matching tracking and precision performance for both these controllers. However from the results of Table. \ref{Tabperformance}, it is clear that CgLp with GSORE outperforms the other. The role of higher order harmonics on performance needs to studied for accurate analysis of these reset elements.

In the case of reset integrator, the introduction of reset results in large increase in tracking error. This is due to the nature of resetting integrator action. Since, the integrator is reset when the error crosses zero, limit cycles are seen in position tracking. This is a well studied problem in literature. This problem is generally solved using a feed-forward controller in parallel which ensures that required steady state output is maintained when the integrator is reset. In the experiments conducted, although a feed-forward controller was used as shown in Fig. \ref{fig:Block}, the feed-forward controller is designed using the inverse of the estimated plant transfer function. Errors in estimation coupled with non-linearity of the leaf flexure stiffness result in an incorrect steady state output from the feed-forward and hence limit cycles. These results further confirm the advantage of using reset action through CgLp for broadband phase compensation rather than the traditional method of resetting the integrator.

From the results of Table. \ref{Tabperformance2}, while tracking performance improvement is seen with use of CgLp, large improvement is not seen from $\gamma = 0.4$ to $\gamma = 0$ with deterioration in tracking performance seen. Interestingly, although the controllers were designed to obtain similar precision performance in terms of gain at high frequencies, improvement is also seen in precision. Additionally, similar to the performance seen in Table. \ref{Tabperformance}, it is noticed that while precision improves with use of CgLp, it does not consistently improve as value of $\gamma$ is lowered, but instead increases again for lower values of $\gamma$.

In the analysis and design of $CgLp$ for both cases, describing function has been used to obtain frequency response behaviour. However, this is only an approximation method. From the results obtained, it can be positively said that this approximation is useful for design. However, from the seen deterioration of results from $\gamma = 0.4$ to $\gamma = 0$ in first case and also deviation of precision performance (although resulting in improvement) in second case, it can be said that more accurate methods than describing function are needed. Since resetting action results in higher order harmonics, these need to be considered to get a more accurate representation of system in frequency domain.\color{black}

\section{Conclusion and Future Work}
\label{Concl}

Industrial workhorse PID is limited by linear controller limitations which can only be overcome by nonlinear controllers. Reset is one such controller which lends itself to standard loop shaping techniques through describing function analysis. While most works in reset have focussed on reduced phase lag of reset filters, this paper has presented a more detailed analysis of reset elements in frequency domain. This knowledge has been used to develop the novel 'Constant in gain Lead in phase' (CgLp) element which is capable of providing broadband phase compensation which had not been explored in literature. 

CgLp-PID controller where the additional phase lead provided by CgLp can be used to improve performance metrics is explained in detail. This concept is tested on one of the DOFs of a precision planar positioning stage and the results validate the improvement expected from theoretical analysis.

However, it is also noted that while all controllers designed for first set with $\gamma < 1$ outperformed the linear controller ($\gamma = 1$), performance slightly deteriorated at smaller values of $\gamma$. Similarly deviation in precision performance was noted. While describing function analysis is accurate enough for understanding and preliminary design analysis of CgLp, alternative methods which take the higher order harmonics introduced by reset into consideration are needed to better explain results. Such a tool would also help in better design of reset elements including CgLp.

Also since this paper presents preliminary performance comparison and validation using CgLp, design using GFORE and GSORE (with $\beta_r = 1$) has been tested. Further, frequencies $\omega_r$ and $\omega_f$ of CgLp are heuristically chosen using rules of thumb. However considering the presence of higher order harmonics, the choice of these values will play a significant role in determining performance and this needs to be investigated further. Additionally, if CgLp is designed using GSORE, the value of $\beta_r$ can be used to shape the phase of open loop further. In summary, tuning of CgLp needs to be investigated further to obtain best possible performance.

\bibliographystyle{IEEEtran}
\bibliography{ref}
\end{document}